\documentclass[preprint,12pt]{elsarticle}

\usepackage{amssymb}

\usepackage{setspace}

\journal{ArXiv}

\begin{document}

\begin{frontmatter}



\title{Density Functional study on the transesterification of triacetin assisted by cooperative weak interactions via a gold heterogeneous catalyst: Insights into Biodiesel production mechanisms}


\author[UPCHIAPAS]{Cornelio Delesma}
\author[UPCHIAPAS]{Roger Castillo}
\author[UPCHIAPAS_Mec]{P.Y. Sevilla-Camacho}
\author[IER]{P.J. Sebastian}
\author[IER,CONACYT]{Jes\'us Mu\~{n}iz}

\address[UPCHIAPAS]{Cuerpo Acad\'emico de Energ\'ia y Sustentabilidad, Universidad Polit\'ecnica de Chiapas, Carretera Tuxtla-Villaflores km 1 + 500, Suchiapa, Chiapas, CP 29150, Mexico}

\address[IER]{Instituto de Energ\'ias Renovables, Universidad Nacional Aut\'onoma de M\'exico, Priv. Xochicalco s/n, 
              Col.  Centro, Temixco, Morelos. CP 62580, Mexico}
              
\address[CONACYT]{CONACYT-Universidad Nacional Aut\'onoma de M\'exico, Priv. Xochicalco s/n,
              Col.  Centro, Temixco, Morelos. CP 62580, Mexico} 
             
\address[UPCHIAPAS_Mec]{Programa Acad\'emico de Ingenier\'ia Mecatr\'onica, Universidad Polit\'ecnica de Chiapas, Carretera Tuxtla-Villaflores km 1 + 500, Suchiapa, Chiapas, CP 29150, Mexico}

\begin{abstract}

A Density Functional study predicting a heterogeneous-catalyzed reaction giving rise to biodiesel, was performed. Triacetin was used as a model of triglyceride in the presence of an Au(111) surface as the heterogeneous catalyst substrate. Explicit methoxy molecules were implemented as an alcohol solvent to understand the reaction trajectory along the well known three-step transesterification process. It was found that the reactants and products in the three-step processes are adsorbed to the Au substrate through non-covalent interactions of the electrostatic-type, which are also mediated by a van der Waals attraction. The density of states indicate that the electronic structure nature of Au is preserved after the interaction with the organic moieties. This may be addressed to an enhanced stability of the Au(111) catalyst through the overall reaction. A charge transfer analysis also reveals that the Au surface oxidation is also responsible for the transesterification of triacetin and evidences that molecular gold plays an important role in this catalytic process. Such results may provide fundamental insights into the design of heterogeneous catalysts for biodiesel production.

\end{abstract}


\end{frontmatter}

\section{Introduction}
\label{Introd}

Due to the high energy demand and continuous increase on fossil fuels consumption and its eventual depletion, it is necessary to increase renewable energy use. Biodiesel production can be a reliable solution since it may be obtained from a sustainable origin, such as vegetable oil, fat animal and microalgae.
The conversion of vegetable oils as $Jatropha$ $curcas$\cite{Zabeti:09}, soybean\cite{Li:06} or sunflower\cite{Arzamendi:08}, and other sources as recycled fried oils into biodiesel, may be a sustainable alternative to produce biofuels. The profitability of biodiesel production relies on its use in a mixture with comercial diesel with an acceptable performance and no engine modifications. Biodiesel also gives an improved combustion due to the presence of oxygen and reduces toxicity to the environment. In addition, its biodegradability reduces the greenhouse effect due to a closed CO$_2$ cycle and triggers sustainable development of rural economies. 


Vegetable oils and animal fats are formed by triglycerides that comprise esters of glycerol with three chains of aliphatic or olefinic free fatty acids with 12-24 carbon atoms. Transesterification is the most common technique to obtain biodiesel, where triglycerides react with a low molecular weight alcohol. This is usually performed with a homogeneous catalyst and the result is a mixture of fatty acid alkyl esters (biodiesel) and glycerol.

Despite the available techniques to produce biodiesel such as hydrolysis\cite{Demirbas:03}, pyrolisis\cite{Demirbas:03}, supercritical alcohol transesterification\cite{Saka:01} or base-catalyzed transesterification\cite{Dorado:02}, acid-catalyzed transeserification\cite{Lotero:05} and enzyme-catalyzed transesterification\cite{Du:04}, homogeneous catalysis is the most popular methodology with the aid of alkaline catalysts such as KOH or NaOH and methanol or ethanol as the alcohol\cite{Zhang:03}. The use of homogeneous catalysts presents acceptable conversion rates, but the process is not as profitable as that obtained with conventional diesel, since the homogeneous catalyst is consumed. The use of high-quality feedstock (virgin or refined vegetable oil) is elementary since water and free fatty acids provoke side effects such as saponification or hydrolysis. As a consequence, the use of a heterogeneous catalyst would reduce the production cost, because a solid is recycled during the production mechanism\cite{Zabeti:09} and esterification and transesterification may be yielded. On laboratory scale, a variety of basic solids including modified zeolites, hydrotalcites, alkaline metal supports have been proven in transesterification process with the aim to solve and simplify biodiesel production \cite{Ebiura2005, Arzamendi2007}.

The stoichiometric reaction requires 1 mole of triglyceride and 3 moles of methanol to form 3 moles of mono-alkyl ester and 1 crude glycerol\cite{Lopez2005,Endalew2011}. (see Fig.\ref{Scheme_reaction}). The homogeneous catalyst provides a faster reaction than the heterogeneous catalyst. However, in  a homogeneous catalyst process, it is required more water to transfer catalysts from the organic phase to an aqueous phase. Therefore, the separation is more expensive for the final product.


It has recently been studied the effectiveness of solid acids catalysts such as Amberlyst$^{\textregistered}$ 15, Nafion$^{\textregistered}$ NR50, support phosphorus acid, sulfated zirconia, zeolites H$\beta$, and ETS-10 (H) \cite{Taramasso1983,Kuznicki1989}. Furthermore, solid base catalysts such as  MgO and ETS-10 (Na, K)\cite{Lopez2005} can catalyze the transesterification reaction. However, the only activation energy known at experimental level corresponds to the first phase triglyceride-diglyceride reaction, ($\sim$12 kcal/mol) using Nafion$^{\textregistered}$ SAC-13\cite{Lopez2007}, as heterogeneous catalyst.

	On the order hand, it has been found that when molecular gold is considered on a size smaller than 10 nm, it is able to catalyze reactions such as CO oxidation and organic compounds\cite{Guzman2005}, to form C-C bonding\cite{Gonzalez2007}, to hydrogenize and to dehydrogenize with high selectivity\cite{Corma2006}.
Haruta et al\cite{Haruta1987} showed exceptional catalytic activity of Au nanoparticles in CO oxidation with low temperature. Mallat and Baiker\cite{Mallat2004} used noble metals as a heterogeneous catalysts to yield selectivity oxidation in OH alcohol groups to carbonyl, using molecular oxygen as oxidant. Theoretical studies showed that the addition of impurities on a Au(111) surface, allows Au to adsorb H atom such as Pd$_{surf}$Au(111)\cite{Venkatachalam2009}.
Bimetallic nanoparticles such as Au@Ag, also has been showed important technology application in heterogeneous catalysts to biodiesel production\cite{Banerjee2014}. 
Gold as a catalyst is directly related with particle size and possible explanations of the phenomena include  the exceptional low coordination of superficial atoms\cite{Janssens2007}, change of oxidation state and electronic structure of metal with the particles size\cite{Wang2007}, sensitivity to adsorption and dissociation of O$_2$ in different arrangements on the Au surfaces\cite{Bornato2010} and charge transfer between Au and metallic oxide supports.

The theoretical studies regarding the description of the homogenous and heterogeneous transesterification are scarce. For instance, Tapanes et al \cite{Tapanes:08} performed a theoretical and experimental study on biodiesel production from Jatropha curcas oil at the AM1 level. A tetrahedral intermediate was found and it was assigned to the difference between kinetics of methanolysis and ethanolysis and the alkoxide formation step. Hori et al\cite{Hori:07} showed a mechanism of the gas-phase acid-catalysed hydrolysis of methyl acetate in gas phase and in solution, using implicit and explicit molecules. It was also found that the reaction proceeds when an explicit solvent molecule is present, since solvent molecules increase the nucleophilicity of the attacking water molecule that corresponds to a weaker nucleophile than the hydroxide ion and allows cleavage of the -OCH group of acetic acid that is included in the reaction. Fox et al\cite{Fox:04} reported that methanol does not bond to carbonyl carbon to create an intermediate structure due to the absence of explicit solvent molecules. Using Density Functional Theory (DFT) at the B3LYP level\cite{Limpanuparb:10}, a systematic study on the kinetic aspects of methanolysis and hydrolysis of triacetin was performed. An acid-catalyzed methanolysis reaction was considered, where it was found that the central carbon-atom of glycerol is reactive in agreement with NMR data. It was proposed that triacetin requires to react with ester carbonyl to allow transesterification. Another DFT/B3LYP study\cite{Silva:13,Silva:14} was also performed on the alkaline-catalyzed transesterification of pentylic acid triglyceride. The calculated activation energies are in reasonable agreement with those found experimentally. On the other hand, a theoretical study of heterogeneous-catalyzed reactions of triglyceride-related moieties has only been performed in the pioneer work of Mu\~{n}iz et al\cite{Muniz:16b}. The interaction of a sulfated zirconia (SZ) slab as a heterogeneous catalyst with triacetin (as a model of triglyceride), was studied at the B3LYP level. Explicit methanol solvent molecules were used and Gibbs free energies revealed the spontaneity of the three-step reaction mechanism for the transesterification of triacetin and an electronic charge transfer among the catalyst, and the organic moieties was evidenced in the reaction. The QST2 scheme was used to calculate the activation energies through the reaction, which is in close agreement with those found experimentally.


The Au-catalyzed heterogeneous reaction has not been experimentally performed and a previous fundamental understanding is crucial to elucidate new methodologies on clean fuels production. The aim of this work is to predict the stability of the Au(111) surface as a heterogeneous catalyst and to give insights into the mechanisms behind the Au-catalyzed transesterification of triacetin, as a model of a triglyceride. This may open new routes to improve Biodiesel production.

\section{Computational Details}
\label{Comp}

\subsection{DFT Calculations}

\begin{figure*}
     \centering
     \includegraphics[height=7cm]{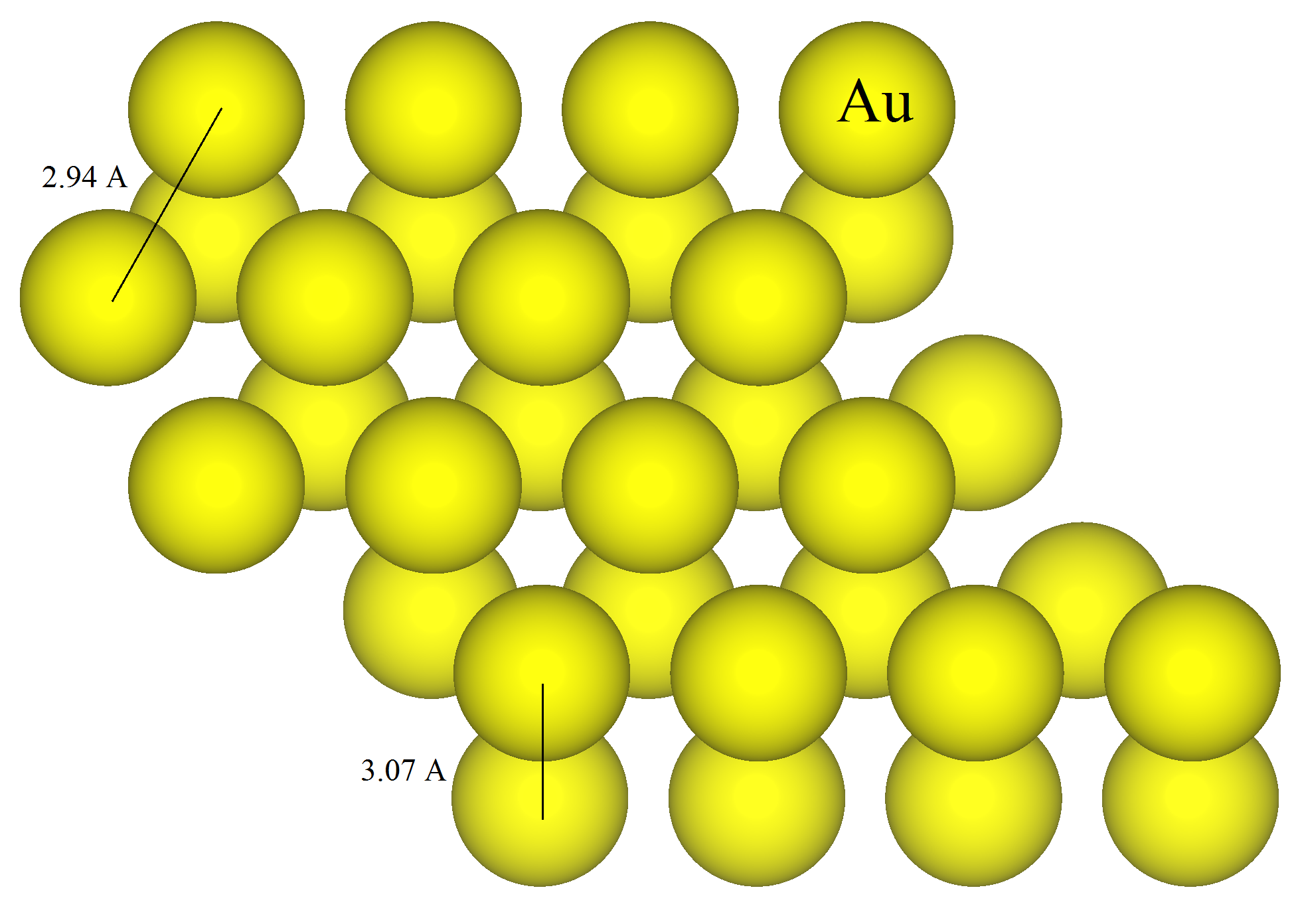}\\
     \centering
    \caption{Structural parameters of the periodic Au(111) surface}
     \label{Fig_Au_surface}
\end{figure*}

All DFT periodic calculations were performed using FHI-aims computational code (Fritz Haber Institute ab initio molecular simulations)\cite{Blum2009}. We employed PBE+vdW\cite{Perdew1996,Tkatchenko2009} functional to describe electron exchange, correlation and London interactions of the van der Waals-type. Relativistic effects were also taken into account at the ZORA (Zeroth-Order Relativistic Approach) level, since it is widely known that they are important for a full description of Au at the molecular level\cite{Pekka1979,Pekka1988,Kaltsoyannis1997,Bartlett1998}. The basis sets \textit{tier 1} was used for the Au atoms, while the \textit{tier 2} basis set was used for the rest of the atoms, as implemented in the FHI-aims code\cite{Blum:09,Havu:09}. Furthermore, the van der Waals (vdW) interactions are important to predict binding forces between Au(111) and the organic molecules, as well as to determine stability for both\cite{Rodriguez2014}. In all calculations, the unit cell has the dimension 20.6 x 20.6 x 30.0\AA, which allows the systems to be confined in a periodic lattice (see Fig.\ref{Fig_Au_surface}). Considering this cell size, the $z$ direction presents a vacuum separation that allows to model the layer structure. It has been shown that this disposition avoids unphysical interactions with images in the periodic lattice. For the convergence criteria, it was considered $0.01\ eV$\AA$^{-1}$ for the final forces in the geometrical relaxations, $10^{-5}$ electrons for the electron density and $10^{-5}$ eV in the total energy of the systems. A Monkhorst and Pack \cite{Monkhorst:76} grid with 3 x 3 x 1 $k$-point for the primitive cell was used to sample the Brillouin zone.

On the other hand, we analyzed electronic structure properties that may give insights into the catalytic process of the reaction trajectories. In this regard, we predicted the energetic stability of the reactants and products by computing the adsorption energies ($\Delta E_{a}$) in accordance with Equation \ref{Eads}:

\begin{equation}
\Delta E_{a}=E_{T}-E_{organic}-E_{Au},
\label{Eads}
\end{equation}

where $E_{T}$ stands for the total energy of the organic moieties adsorbed onto the Au(111) substrate. $E_{organic}$ is the total energy of the organic moieties during the different reaction steps, and $E_{Au}$ corresponds to the total energy of the gold substrate, both inside the same unit cell.


The trajectories of reaction were computed by using the string method\cite{Weinan2002} and the \textsl{climbing images}\cite{Henkelman2000} with a 0.2 eV of energy as a threshold force for both. For the barrier energy convergence criteria, we used 11 images from the reactants to the  products.



Charge transfer was also studied among the organic moieties and the Au(111) substrate in accordance to the M\"ulliken population analysis \cite{Mulliken:55}. As it is well known, the methodology is highly sensitive to the basis set used. As a consequence, we are interested to provide qualitative trends into the electronic structure behavior of the organic moieties adsorbed on the metallic substrate. The method given by Arellano et al \cite{Arellano:00} was used, where the charge density is partitioned in real space in order to state the interaction and quantitatively verify the charge transfer between the two subsystems under study. In this case, the organic moieties and the Au(111) surface. Further, the charge transfer ($\Delta$Q) is define as:
    
\begin{equation}
\Delta Q=\sum Q_{T}-\sum Q_{organic}-\sum Q_{Au(111)},
\label{DeltaQ}
\end{equation}
where $\sum Q_{T}$ represents the sum of all charges at each of the atoms in the nanocomposite system; $\sum Q_{organic}$ and $\sum Q_{Au(111)}$ are the total sum of the charges at the isolated organic moieties and the Au(111) surface, respectively. In order to complete Arellano's scheme, we also mapped the isosurfaces of the total charge density difference $\rho_{diff}(r)$ for the reactants, products and transition states. In this respect, $\rho_{diff}(r)$ was plotted in the molecular viewer VESTA \cite{Momma:11}, according to Equation \ref{rhodiff}
\begin{equation}
\rho_{diff}(r)=\rho_{organic/Au(111)}(r)-\rho_{organic}(r)-\rho_{Au(111)}(r),
\label{rhodiff}
\end{equation}
where $\rho_{organic/Au(111)}(r)$ refers to the charge density of the organic moieties adsorbed on the Au(111) substrate;  $\rho_{organic}(r)$ is the charge density of the organic moieties  and $\rho_{Au(111)}(r)$ corresponds to the charge density of pristine Au(111) surface, both in the same unit cell.

\section{Results and discussion}
\label{Result}

\subsection{PBE and PBE+vdW calculations}
\label{sec:pbe_vdw}

The three-step mechanism of the transesterification reaction, as suggested by L\'opez et al\cite{Lopez2005,Lopez2007} was studied with electronic structure methodologies in accordance to DFT and using the computational details described above. As it has previously been reported, there are 12 elementary-step\cite{Limpanuparb:10} mechanisms that triacetin may follow in order to achieve transesterification and finalize the process in the conversion to methyl acetate. Nevertheless, to the best of our knowledge, there is no experimental evidence that reveals the kinetics of the last two steps in presence of a heterogeneous catalyst. As a consequence, we explore one of such reaction mechanism as proposed by L\'opez et al\cite{Lopez2005}, and by considering the effect coming from an Au(111) surface, which activates the reaction. We have chosen Au, due to its enhanced catalytic activity, as observed at the molecular level in analogous reactions\cite{Rodriguez2014,Carrasco:14}.

\begin{figure*}
     \centering
     \includegraphics[height=7cm]{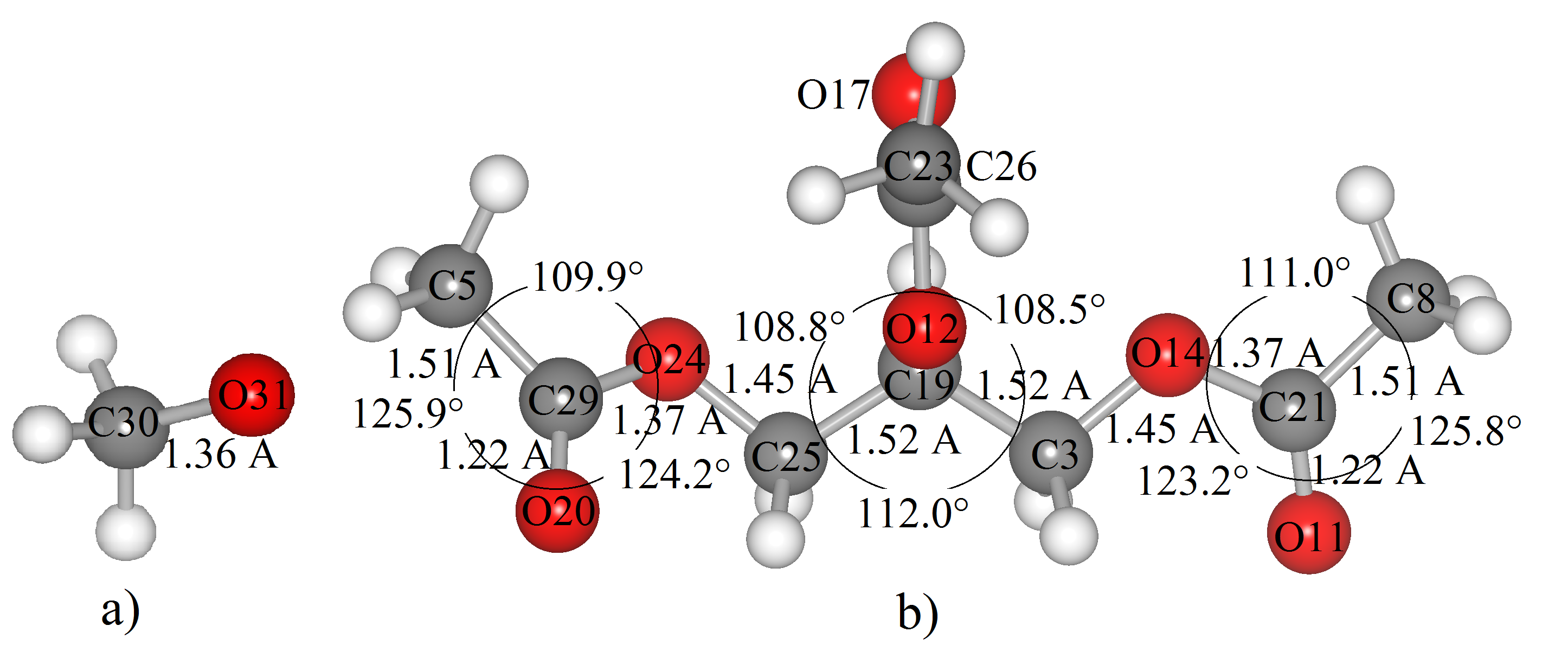}\\
     \centering
    \caption{Optimized structural parameters of  (a) Methoxy, (b) Triacetin}
     \label{Fig_Meth_Triacetin}
\end{figure*}

As previously suggested, no experimental or theoretical results have been reported where Au(111) catalyst acts as a solid support in the production of biodiesel. In this regard, we have simulated the three-step reaction by optimizing the geometry of each of the components in the reaction mechanism. That is, the geometry of the solvent molecule (methoxy), was fully relaxed in accordance to the computational details. The structural parameters are reported in Fig.\ref{Fig_Meth_Triacetin}(a). Triacentin was used as a model of a triglyceride (see Fig.\ref{Fig_Meth_Triacetin}(b)), since all electronic structure properties are finely reproduced\cite{Lopez2007}.

\begin{figure*}
     \centering
     \includegraphics[height=7cm]{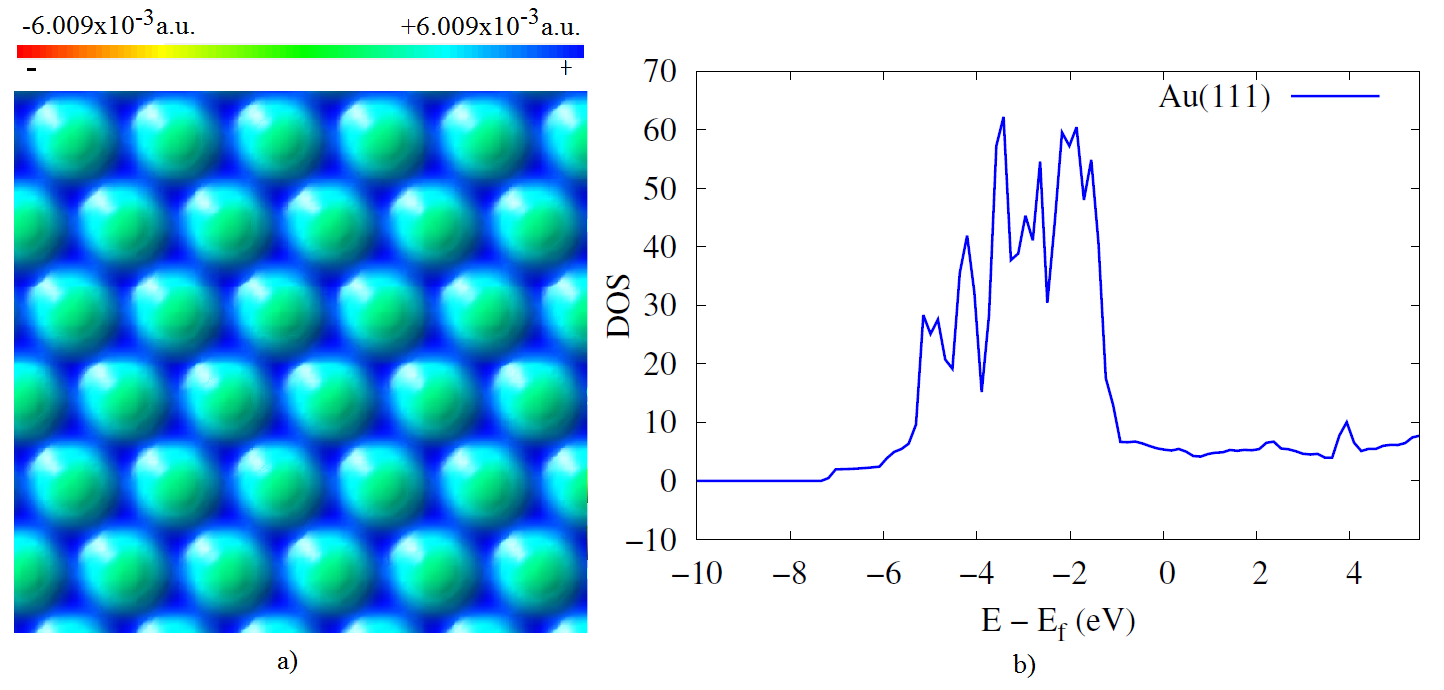}\\
     \centering
    \caption{(a) Molecular electrostatic isosurface of bare Au(111) surface, (b) Density of states (DOS) of bare Au(111) surface (solid line). Note that 0 eV corresponds to the Fermi level}
     \label{Fig_MEP_Au}
\end{figure*}

In order to model the Au(111) catalyst, we used a slab model of Au with two layers and an interplanar distance of 3.07{\AA} and an Au-Au bonding distance of 2.94{\AA}, which corresponds to the distance of one Au atom with respect to its nearest neighbor. Such structural parameters are in agreement with those reported experimentally\cite{Krupski:15}. We only considered the presence of two interatomic layers, since it has been shown\cite{Rodriguez2014} that the catalytic effects on Au(111) surfaces take place on the most external layer of the substrate. Taking the latter into account, we relaxed the metallic surface by fixing the internal layer, and we only allowed the outermost layer to fully relax during the geometry optimization. Furthermore, we mapped the molecular electrostatic potential isosurface (see Fig.\ref{Fig_MEP_Au}(a)) onto the electronic density to have an insight into the acid/base nature of molecular Au(111). A basic behavior was found, which is evidenced from the negative charge distribution located at the Au atoms, and shown as the light yellow regions at the Au atoms.

As a proof on the validity of our simplified model of Au(111) surface, we performed a Density of States (DOS) calculation and the result is depicted in Fig.\ref{Fig_MEP_Au}(b). The DOS reveals the expected metallic nature of Au due to the population of electronic states around the Fermi level, which is in close agreement with the DOS reported for pristine Au(111) surface, as presented by Venkatachalam et al\cite{Venkatachalam2009}. In this model, six-layers of Au atoms were considered to simulate the Au(111) surface, but only the two most external layers were allowed to relax. As a consequence, we found that the inner layers in the Au(111) substrate do not affect the electronic structure nature.

The followed-up three-step reaction on the transesterification of triacetin was simulated with the DFT methodology reported in the Computational Details section. In our scheme, we fully optimized the molecular geometries of the reactants and products in each of the steps. In Table \ref{Table_Ads_Ener}, we present the adsorption energies for those configurations in accordance to Equation \ref{Eads}. In all steps, we found that the adsorption energies range below 100 kcal/mol, which may be addressed to non-covalent interactions of the electrostatic-type. We also performed such calculations at the DFT/PBE level without the inclusion of the vdW dispersive term in order to elucidate if the London-type attractions were also involved in the overall interaction. The adsorption energy calculations computed at the PBE level with no vdW effects are also presented in the values on parenthesis presented in Table \ref{Table_Ads_Ener}. It may be seen that the vdW contribution stabilizes the attraction of the organic moieties on the Au(111) substrate, since a significant energy difference ranging from -26.2 to -44.3 kcal/mol, was found (as shown in Table \ref{Table_Ads_Ener}). This may indicate that the attraction on the surface may strongly be mediated by a vdW interaction. This is also in agreement with the adsorption energy of analogous systems such as organic acids\cite{Rodriguez2014}, where the vdW dispersive forces rules the attraction with the Au(111) surface.

\subsection{Reaction Mechanism}

\begin{figure*}
     \centering
     \includegraphics[height=7cm]{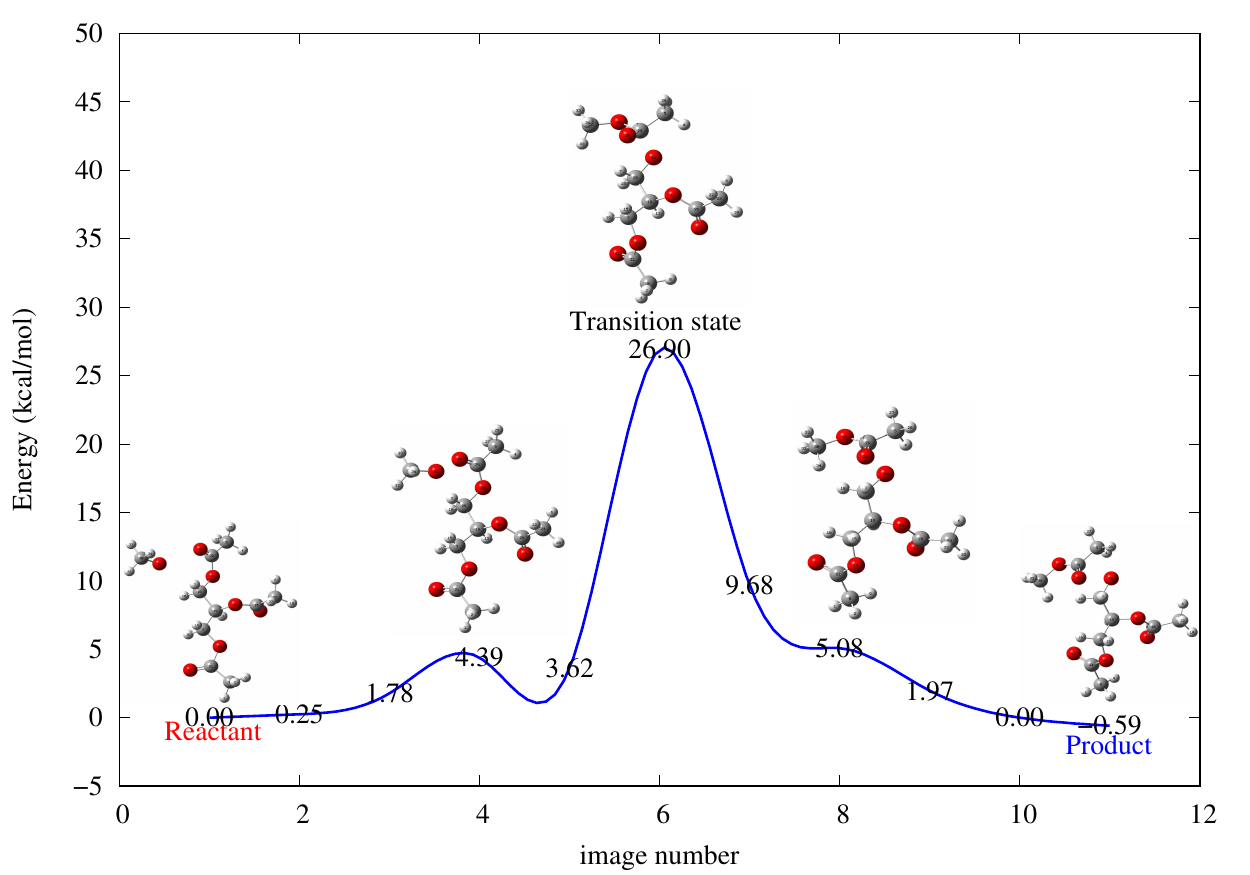}\\
     \centering
    \caption{First step: Trajectory of reaction from triacetin to diacetin in the presence of the Au heterogeneous catalyst. Note that the Au(111) surface has been omitted for clarity}
     \label{Fig_Step_1_RS}
\end{figure*}

\begin{figure*}
     \centering
     \includegraphics[height=7cm]{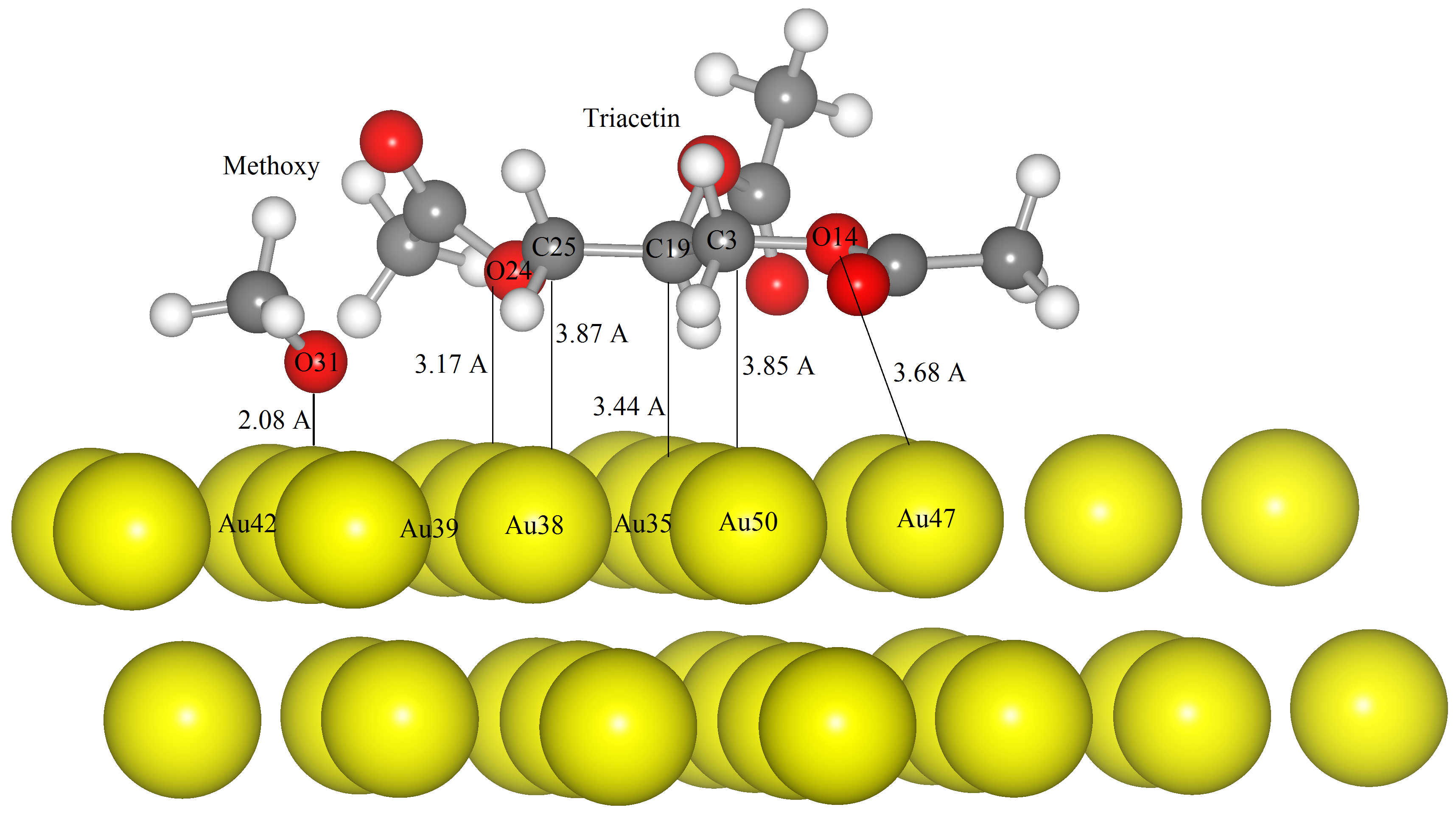}\\
     \centering
    \caption{Selected optimized structural parameters of the reactants at step 1 of the overall reaction}
     \label{Fig_Step_1_image1}
\end{figure*}

The three-step reaction, as proposed by L\'opez et al\cite{Lopez2005} was modeled in accordance to the string method, as listed in the Computational Details. The results for the first step are shown in Fig.\ref{Fig_Step_1_RS}, where the number of different configurations, which we may call 'images', are mapped throughout the reaction trajectory. The first image corresponds to the optimized reactants, where triacetin was relaxed with a methoxy molecule as a model of explicit solvent in the presence of the Au(111) surface. In this image, the methoxy is adsorbed on Au(111) through an Au-O bond length of 2.00{\AA}, as it is shown in Fig.\ref{Fig_Step_1_image1}. On the other hand, triacetin is weakly adsorbed to the Au(111) substrate, where the closest Au-O contact is located at 3.17{\AA} (see Fig.\ref{Fig_Step_1_image1} and Table S1 of Supplementary Information SI). The overall adsorption energy amounts to -62.59 kcal/mol, which may be assessed to a weak attraction of the electrostatic-type.

It was found that the strong adsorption of the methoxy on the Au(111) surface, increases the C-O bond lengths up to 0.6{\AA}, which may be due to the strong Au-Au interaction in the substrate with the molecule that induces the breakage of the initial covalent bond observed on the methoxy. The modified parameters of triacetin adsorbed on Au(111) with respect to the isolated molecule are reported in Table S2 of the SI, where no significant changes were reported, indicating that triacetin remains structurally unaltered.

This initial image was considered as the frame of reference (0.0 kcal/mol) with regard to the rest of the geometric configurations. The geometry of the combined system triacetin/methoxy is virtually unaltered with respect to (wrt) the first image and only slight deviations at the central bond angles are reported (as presented in Table S2 and Fig.S1 of SI). Larger deviations at the bond angles for image 3 in triacetin are reported (see Table S3 and Fig. S2) wrt image 2. This results in a slight increase in the total energy (1.78 kcal/mol). The variations on the bond angles for image 4 wrt image 3 range up to 2.0$^{\circ}$ (see Table S4 and Fig. S3 of SI), which destabilizes the total energy of the system to about 4.5 kcal/mol, which is readily interpreted as the first transition state (TS1) for this reaction (see Fig.\ref{Fig_Step_1_RS}).

\begin{figure*}
     \centering
     \includegraphics[height=12cm]{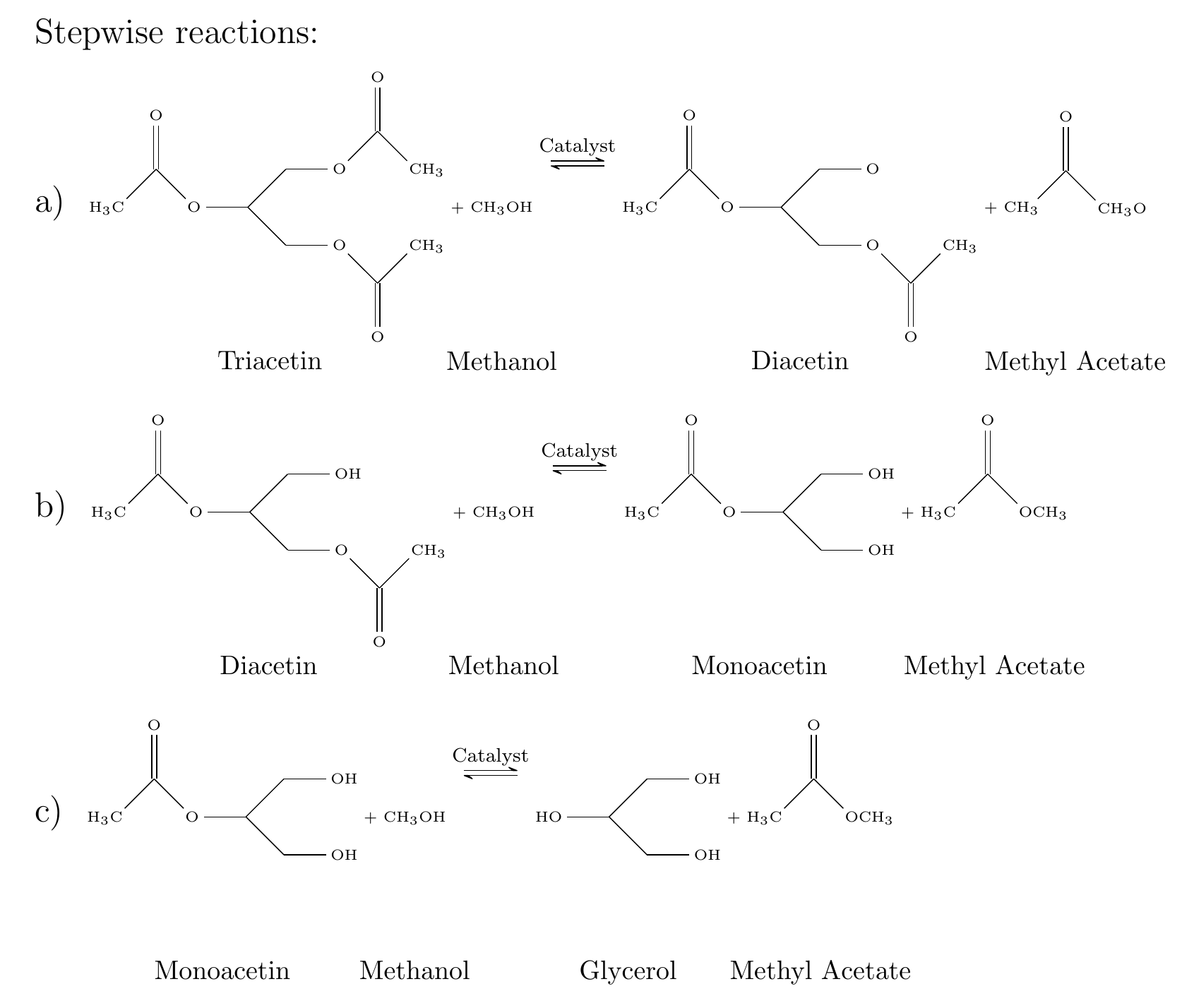}\\
     \centering
    \caption{Three-step overall reaction of triacetin transesterification in the presence of a solid heterogeneous catalyst}
     \label{Scheme_reaction}
\end{figure*}

\begin{figure*}
\begin{minipage}[t]{0.30\textwidth}
     \includegraphics[height=4.5cm]{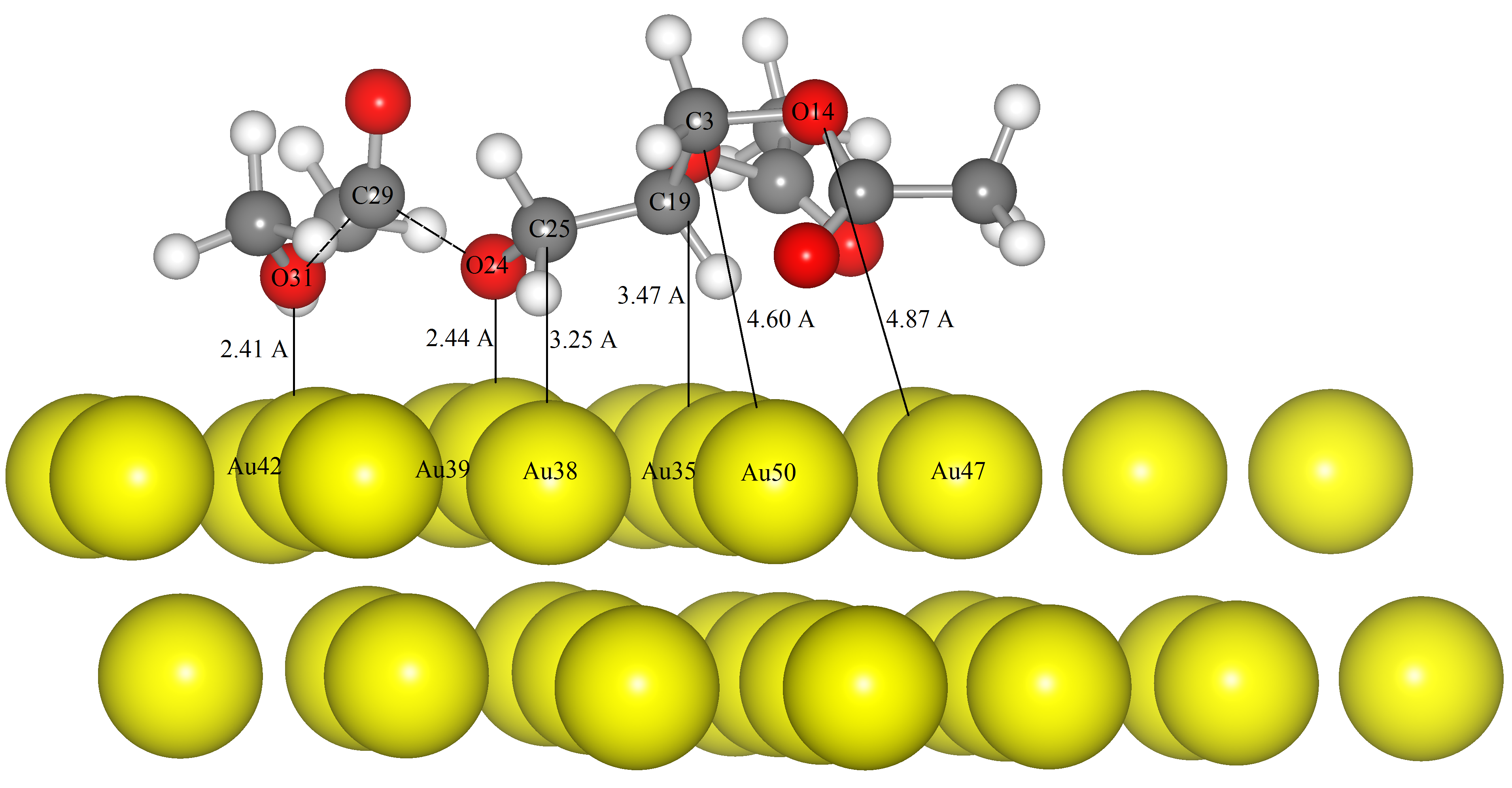}
     \begin{center}
	a)
\end{center}
\end{minipage}
\hfill
\begin{minipage}[t]{0.30\textwidth}
\includegraphics[height=4.5cm]{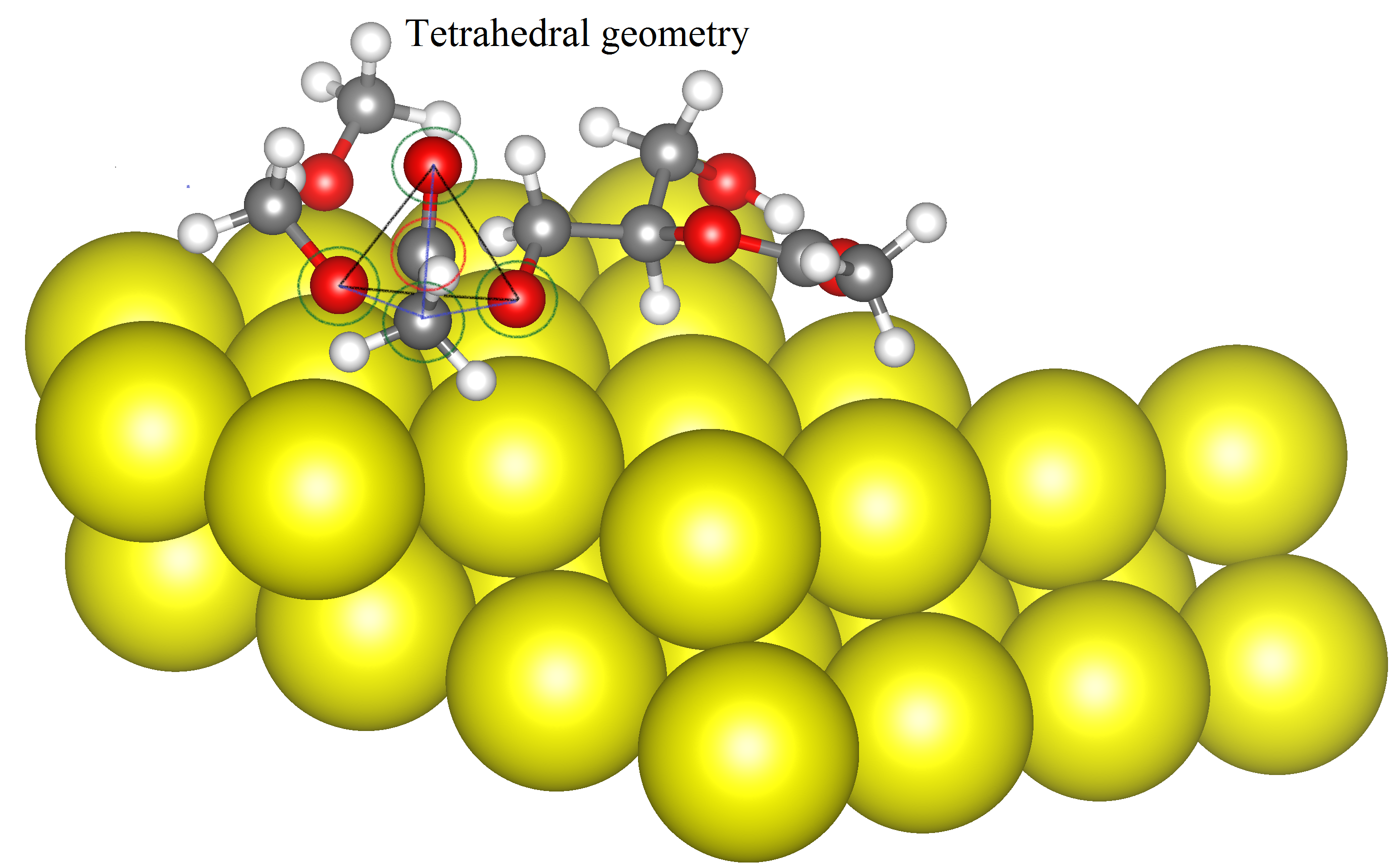}
\begin{center}
b)
\end{center}
\end{minipage}
\caption{(a) Transition State TS2 of the first step in the overall reaction, (b) Tetrahedral intermediate (TI) located at a transition state during the second step of the overall reation}
\label{Fig_ts2_step1}
\end{figure*}

Furthermore, at image 6, as depicted in Fig.\ref{Fig_Step_1_RS}, the spontaneous breaking of the C$_2$H$_3$O group of triacetin was observed, which may be readily addressed to the rising of the transesterification process. This is also related to the origin of a second transition state (TS2), where the C$_2$H$_3$O group reacts with methoxy to form methyl acetate (Biodiesel) and the side-formation of diacetin is found, as depicted in Fig.\ref{Fig_Step_1_RS}. This is in agreement with the transesterification process depicted in Fig.\ref{Scheme_reaction}. It was also found that the activation potential ($E_a$), which is defined as $E_{a}=E_{TS} - E_0$ (where $E_{TS}$ is the total energy of the transition state and $E_0$ is the ground state energy of the reactants), amounts to 26.90 kcal/mol. This value may also be comparable with the activation potential found for the analogous Nafion$^{\textregistered}$ SAC-13 heterogeneous-catalyzed triacetin to diacetin reaction, observed by L\'opez et al\cite{Lopez2005}. The closest contact between the methyl acetate and Au(111) is through the Au42-O31 bond length at  2.41{\AA} (see Fig.\ref{Fig_ts2_step1} (a)), which may be addressed to a non-covalent interaction with a certain degree of dispersive attraction of the vdW-type.

\begin{figure*}
     \centering
     \includegraphics[height=7cm]{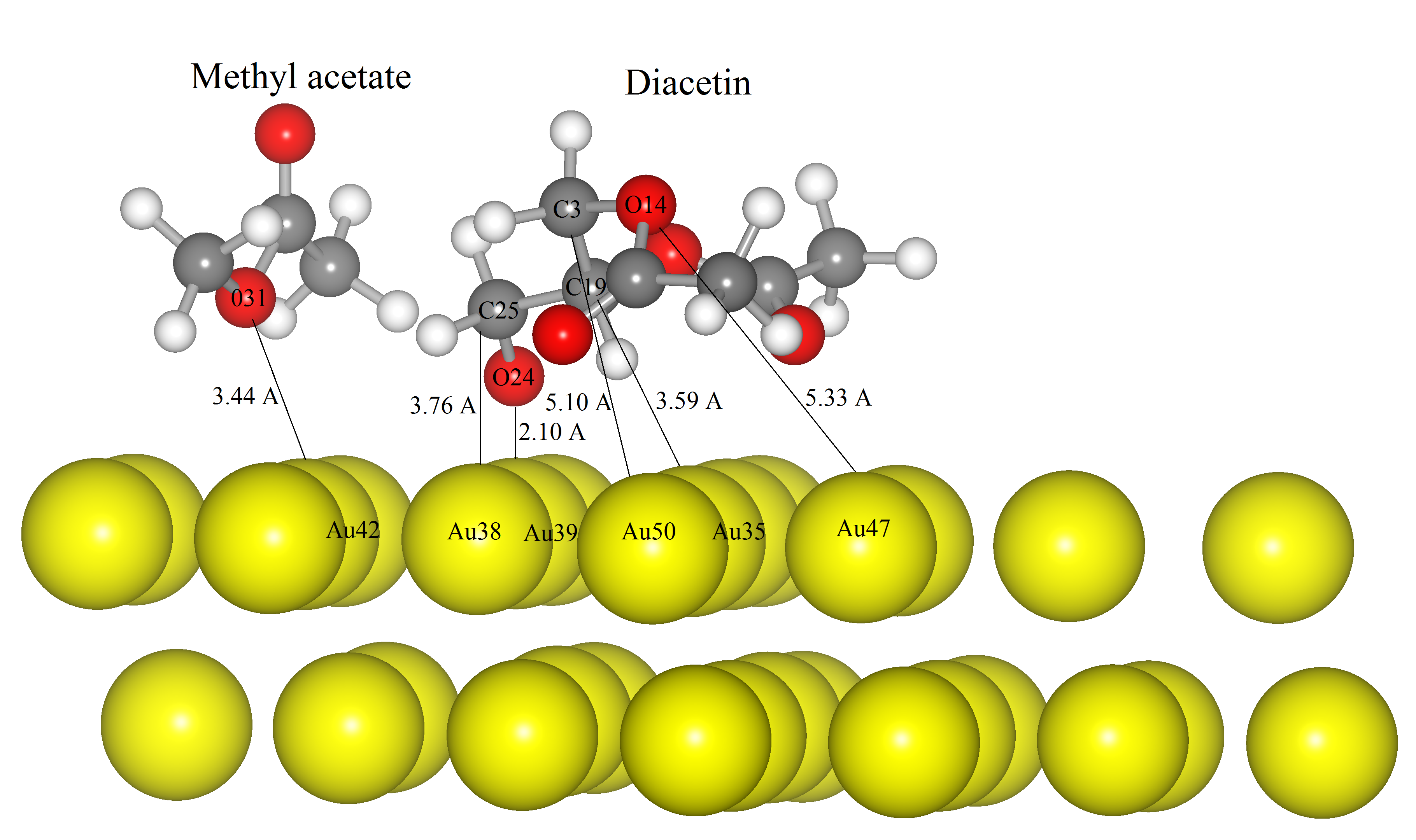}\\
     \centering
    \caption{Final product methyl acetate and triacetin at the first step in the overall reaction.}
     \label{Fig_fin_pro_step1}
\end{figure*}

The Au(111) surface keeps the original symmetry with only slight variations at the Au-Au contacts, which may be considered to be negligible. This may be related to a high stability in the heterogeneous catalyst that may be ascribed to its reusability through the different reaction steps. The formation of diacetin on the Au(111) substrate reveals a weak interaction that is supported by Au-O contacts and  Au-C contacts at 3.25{\AA}. Such structural changes are presented in Table \ref{Table_ts2_step1}. Furthermore, the overall structural changes are reported in Table S5. It was found that the transition state TS2 is basically raised with the formation of a C5-O29-O24 bond angle that corresponds to the central angle of methyl acetate (see Fig. S4). The evolution continues from image 7 to 12 (see Fig.\ref{Fig_Step_1_RS}). As depicted in Fig.\ref{Fig_fin_pro_step1}, the methyl acetate is weakly adsorbed on the Au(111) substrate with the closest Au-O contact at 3.44{\AA}. This makes it suitable to be separated from the heterogeneous catalyst. The thermodynamical energy is defined by the following relation: $E_{Th}=E_{f}-E_{i}$, where $E_{Th}$ is the thermodynamical energy of the system, $E_{f}$ is the energy of the final state (products) and $E_{i}$ is the energy of the initial state of the system (reactants). In the first step of the reaction, it corresponds to -0.59 kcal/mol, which may be interpreted as a spontaneous exothermal reaction, addressing its feasibility.

\begin{figure*}
     \centering
     \includegraphics[height=7cm]{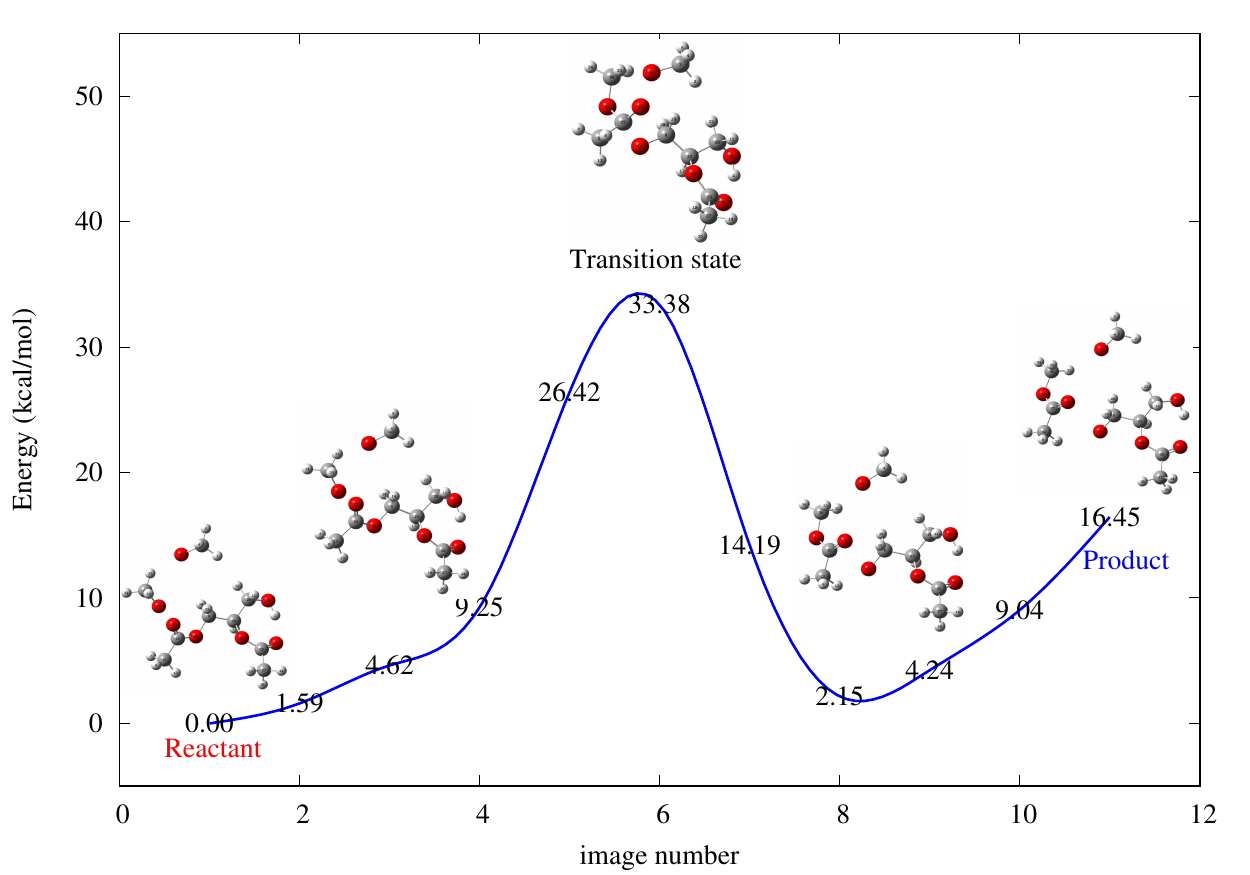}\\
     \centering
    \caption{Second step: Trajectory of reaction from diacetin to monoacetin in the presence of the Au heterogeneous catalyst. Note that the Au(111) surface has been omitted for clarity}
     \label{Fig_Step_2_RS}
\end{figure*}

In order to model the second step of the overall reaction (see Fig.\ref{Scheme_reaction}(b)), we computed the trajectory of the reaction using the same methodology. The resulting full reaction path is depicted in Fig.\ref{Fig_Step_2_RS}. The first image corresponds to the optimized structure of diacetin with a methoxy molecule, also used as an explicit solvent molecule. The combined system was fully relaxed in the presence of the Au(111) surface. At this configuration, the closest oxygen atom of the methoxy is located at 2.10{\AA} above the Au(111) surface (see Fig. S5), corresponding to the Au42-O24 bond length, while the shortest bonding of diacetin to the Au surface is located at 2.23{\AA}. The adsorption energy for this image corresponds to -95.39 kcal/mol, which may also be addressed to a non-covalent attraction with a vdW dispersive character, just as it was already reported in the first step. It is important to highlight that two methoxy solvent molecules were implemented in the calculation with the string methodology to describe all interactions present in this trajectory. A close-up of  image 1 is shown in Fig. S6. Note that the geometry reported was obtained with the presence of the Au(111) catalyst, and it was omitted in the Figure for clarity. During the first 5 images of the reaction trajectory, the methoxy solvent molecules present no significant changes. Nevertheless, the increasing in total energy is produced due to the structural changes in diacetin, which are basically located at the bond angles formed by the O-C-C at the center of the molecule. During the evolution of the diacetin molecule, it was found that a first transition state (TS1) is located at image 3 with a small activation energy of 4.62 kcal/mol. The rising of a second transition state (TS2) is located at image 6, as depicted in Fig.\ref{Fig_Step_2_RS}. At this step, the C$_2$H$_3$O fragment is detached from diacetin and reacts with one of the methoxy molecules to give rise to methyl acetate. Due to the loss of this fragment, the rising of monoacetin is formed, as suggested in Fig. S7. The closest contacts of the organic moieties with respect to the Au(111) surface are depicted in Fig. S7. The main structural parameters are also reported in the close-up of Fig. S8 of SI.

It is important to highlight that the tetrahedral intermediate (TI) formation is also observed at this step (see Fig.\ref{Fig_ts2_step1}(b)). The rising of TI is a characteristic signature found at analogous homogeneous-catalyzed reactions\cite{Limpanuparb:10}. Such geometry rises after the interaction of C25 of methyl acetate and O10 of monoacetin. The activation energy for TS2 amounts to 33.48 kcal/mol, which is slightly larger than that found in the first step, but it does not compromise the feasibility of the reaction. The evolution of the trajectory is depicted in Fig.\ref{Fig_Step_2_RS} after the second transition state is reached, and no significant changes are reported. The total evolution of the energy in the organic system supported on the Au(111) surface is decreased and it is stabilized at image 11, which corresponds to the products at the second step of the reaction (see Fig. S9). The closest methoxy/Au(111) contact was located at 2.10{\AA}, while the closest methyl acetate/Au(111) surface contact is located at 3.81{\AA}, which may be ascribed to a weak interaction of the electrostatic-type, also mediated by a vdW dispersive contribution (with an adsorption energy of -99.84 kcal/mol). The monoacetin/Au(111) contacts range from 2.10{\AA} to 3.71{\AA} and it is weakly adsorbed on the substrate. It was also found that during the evolution of both reaction trajectories, the Au(111) surface undergoes a slight change on the Au-Au contacts up to 0.5{\AA}, which is not directly compromising the trajectory of triacetin, but it may be considered as a cooperative agent driving the transesterification catalytic reaction already evidenced through both trajectories. The thermodynamical energy for the second step is +16.45 kcal/mol, which corresponds to an endothermal reaction that may require heat in order to be achieved.

\begin{figure*}
     \centering
     \includegraphics[height=7cm]{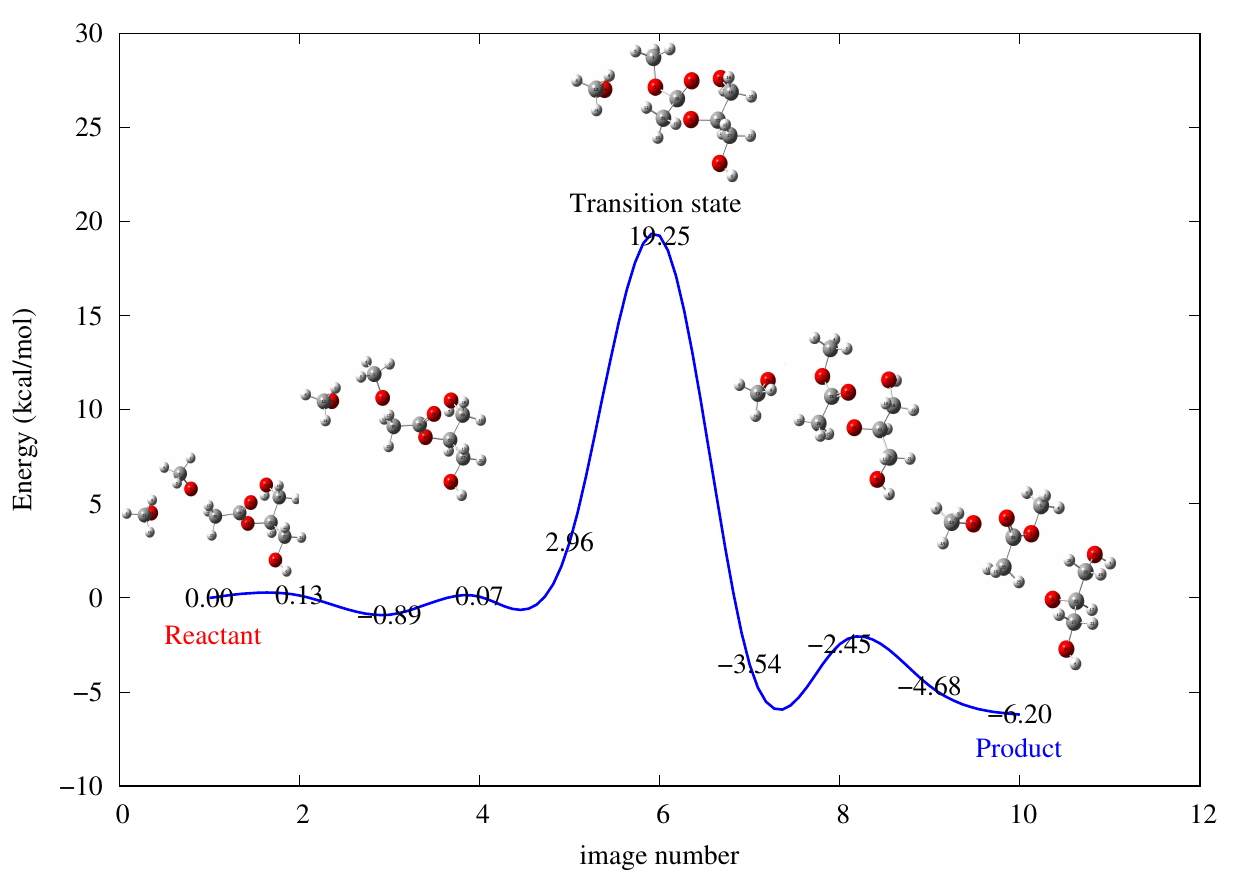}\\
     \centering
    \caption{Third step: Trajectory of reaction from monoacetin to glycerol in the presence of the Au heterogeneous catalyst. Note that the Au(111) surface has been omitted for clarity}
     \label{Fig_Step_3_RS}
\end{figure*}

The last step in the reaction was also modeled with the same methodology described above. The complete reaction trajectory is presented in Fig.\ref{Fig_Step_3_RS}, showing that the third step of the overall reaction involves monoacetin and a methoxy solvent to produce methyl acetate and glycerol as byproduct. As previously performed in the former step of the overall reaction, we used 2 methoxy units as explicit solvent molecules. The geometry at image 1 in Fig. S10 reveals that the reactants in both methoxy solvent molecules are attached to the Au surface with Au-O bonding pairs located at 2.08{\AA} and 2.26{\AA}, while the shortest bond length of monoacetin to the Au substrate is located at 2.74{\AA}. Selected bond lengths and angles are listed in Table S6. The parameters are also characteristic of non-covalent attraction of the electrostatic-type. The trajectory through the reaction shows 2 transition steps that may be identified with small activation potentials of 0.13 and 0.07 kcal/mol, which are identified as transition states TS1 and TS2, respectively. The structural parameters through images 2 to 5 are not significantly altered and the rising of transition state 3 (TS3) was found with an activation potential of 19.25 kcal/mol, which is also comparable to the activation barriers found in the previous reaction steps. At TS3, the detachment, i.e., transesterification of a C$_2$H$_3$O$_2$ group in the monoacetin is observed, and the rising of the methyl acetate (biodiesel) is obtained (a close-up of this configuration is presented in Fig. S11 of SI). The full reaction coordinate is completed from images 7 to the final image 10, which corresponds to the products. The geometrical representation is depicted in Fig. S12: The methoxy remains tightly bounded to the Au surface with a bond length of 2.09{\AA} at the shortest Au-O contact. Methyl acetate and glycerol are weakly adsorbed with Au-O contacts at 3.08{\AA} and 2.12{\AA} from the Au(111) surface plane, respectively.

The adsorption energy for this image is -87.27 kcal/mol, which belongs to the same magnitude found in the former 2 reaction steps. The main structural parameters and a close-up of the organic moieties in this image are reported in Table S7 and Fig. S13 of SI. For this last reaction step, a thermodynamical energy of -6.20 kcal/mol was computed, which corresponds to an exothermal reaction that may spontaneously take place. Note that the Au-Au bonding length parameters present slight elongations no longer than 0.5{\AA} at the regions of contact with the organic moieties. This indicates that the presence of the catalyst does not rule the transesterification process, but its effect on the reaction can not be neglected.

\subsection{Charge Density difference isosurfaces and Mulliken analysis}

\begin{figure*}
\begin{minipage}[t]{0.30\textwidth}
\includegraphics[height=2.5cm]{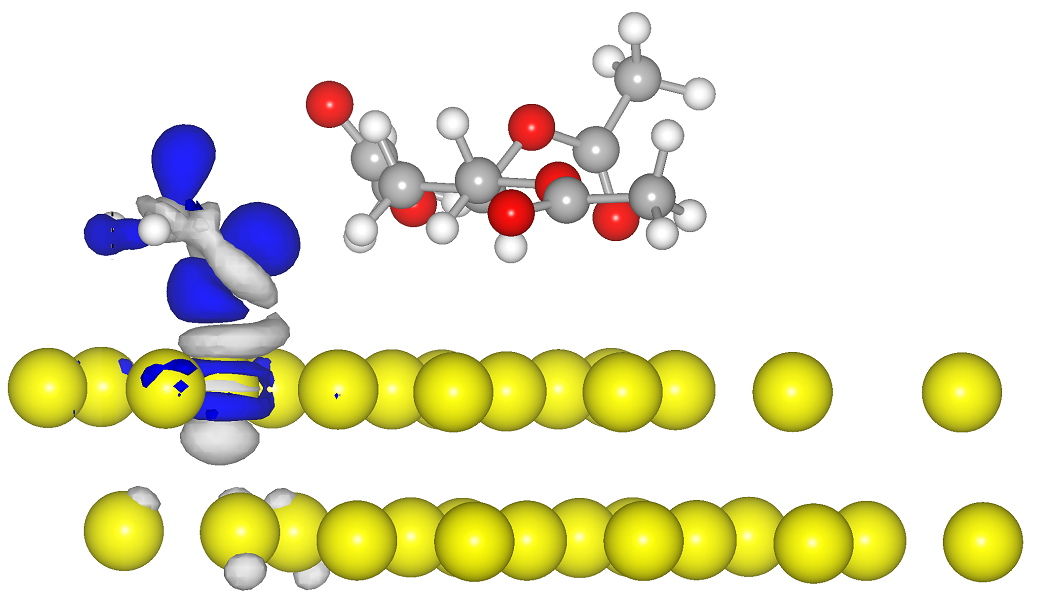} 
\begin{center}
a)
\end{center}
\end{minipage}
\hfill
\begin{minipage}[t]{0.30\textwidth}
\includegraphics[height=2.5cm]{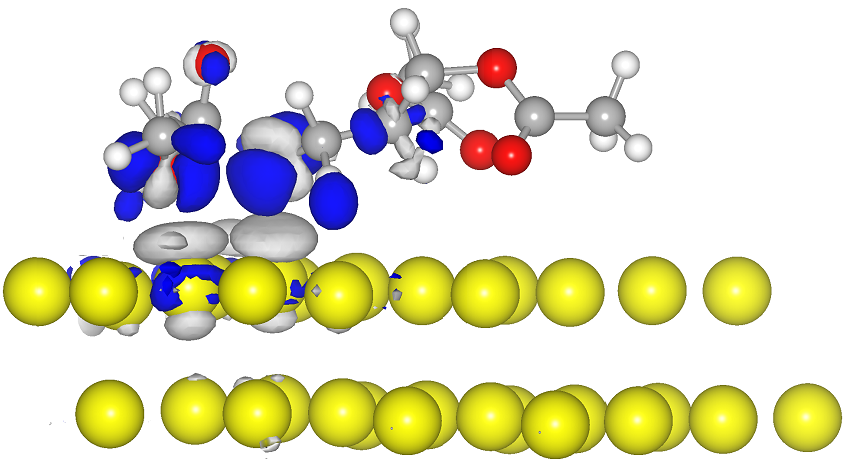} 
\begin{center}
b)
\end{center}
\end{minipage}
\hfill
\begin{minipage}[t]{0.30\textwidth}
\includegraphics[height=2.5cm]{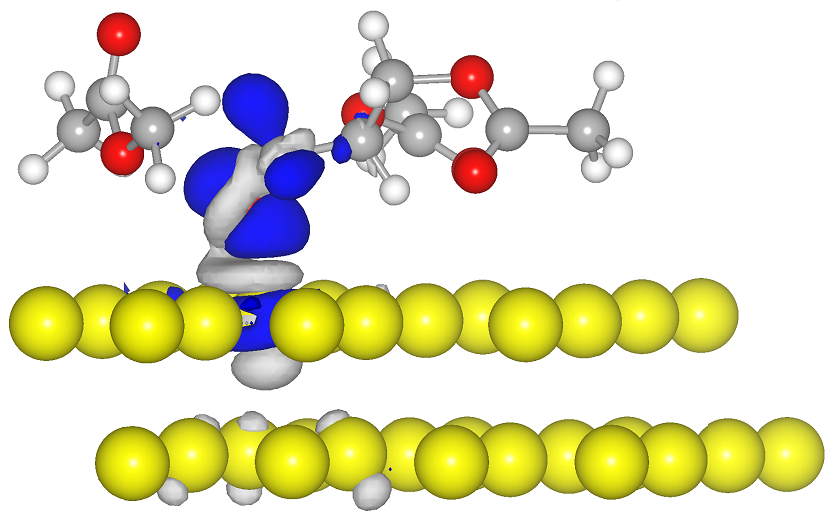} 
\begin{center}
c)
\end{center}
\end{minipage}
\caption{Charge density difference isosurfaces for (a) the reactants, (b) transition state TS2 and (c) products for the first step of the overall reaction}
\label{Fig_Rhos_Step1}
\end{figure*}

In order to understand the mechanism behind charge transfer among the organic moieties and the Au substrate, we performed the charge difference ($\Delta$Q) analysis scheme, as proposed by Wen et al\cite{Wen:12} and Mu\~{n}iz et al\cite{Muniz:16a}; where $\Delta$Q is given in Eq.\ref{DeltaQ}  of the computational details section. The results are given for the 3 steps of the reaction in Table S8, for each of the reactants, products and transition states with larger activation energies. The same tendency is present on each of the steps and a clear depletion of electronic charge was found at the Au centers, which indicates that a charge transfer is present for each of the steps, and it is coming from the Au(111) surface. Taking the latter into account, we performed a charge density difference analysis ( $\rho_{diff}(r)$ ) as suggested by Eq.\ref{rhodiff}. Such equation was considered to plot the isosurfaces of density difference. The gray regions correspond to zones where the electronic charge was transferred to, while the blue regions represent those zones where the charge was depleted. The $\rho_{diff}(r)$ isosurfaces for the main images of the reaction trajectory in step 1 are presented in Fig.\ref{Fig_Rhos_Step1}. In all images, a charge transfer coming from the Au centers is located at the interplanar region between the Au(111) surface and the organic moieties. This is in agreement with the $\Delta$Q Mulliken values that were found for those systems. Such behavior is enhanced at the second transition state (TS2) found for this step, where a larger amount of charge is transferred to the region above the Au surface (see Fig.\ref{Fig_Rhos_Step1}(b)). Consequently, a larger amount of charge re-located among the solvent, the detached triacetin and the presence of the Au catalyst, may be assigned as one of the agents giving rise to the transesterification of triacetin observed during the reaction trajectory for this step.

The same pattern was also found in steps 2 and 3 (see Fig S14 and S15). The transition state 2 (TS2) and TS3 in steps 2 and 3, respectively, present a dramatic charge transfer from the organic moieties to the Au surface, which is also strenghthened from the charge transfer of Au to the same region. The charge relocation assessed in the three steps may be directly interpreted as one of the driving forces that lead to transesterification and highlights that the presence of the Au substrate plays a fundamental role in order to achieve it.

\subsection{Molecular Electrostatic Potential (MEP) surfaces and Density of States (DOS)}

In order to study the Au(111) electronic structure modification with the presence of the organic moieties, a MEP isosurface was mapped onto the Au(111) substrate with the optimized reactants of step 1. It was found that the basic nature of bare Au(111) is changed to an acidic character instead. This is due to the overall charge transferred from the organic moieties to the Au surface (see Fig. S16). The same tendency was found for the products. Such behavior was also verified in the last two steps of the overall reaction coordinate, as shown in Fig. S17 and S18.

On the other hand, the enhancement of the acidic character of the Au(111) substrate is also evidenced on the DOS (see Fig. S19)  for the products of step 1, where the density of states below the Fermi energy are strengthened due to the charge transfer at the intermediate region between the Au(111) surface and the organic moieties. Note that the modifications of the DOS is caused by the change transfer, but it does not change the metallic character of the substrate, which may be addressed to the stability of the Au(111) substrate along the reaction trajectory. Besides, Fig. S20 and S21 of SI show an  equivalent behavior, for the products of step 2 and 3, respectively. Consequently, the Au(111) surface may be considered as a potential candidate to be implemented as heterogeneous catalyst that may potentially aid biodiesel production.

\section{Conclusions}

A DFT study on the reaction mechanism leading to transesterification through an Au heterogeneous-catalyzed reaction was performed by the first time. The reaction mechanisms for the followed-up three-step reaction starting from triacetin, reveals the rising of two transition states at each of the steps. The activation energies in the three steps are in close agreement with that found experimentally for a heterogeneous-catalyzed reaction, highlighting the feasibility of the Au(111) substrate to be incorporated in the transesterification of triglycerides. It was found that the Au support does not significantly change the electronic structure nature along the reaction trajectory, which guarantees catalyst stability along the reactive process. The morphology of the catalyst remains also constant during the three-step reaction, and a charge transfer coming from the Au surface indicates that the transesterification process is ruled by a charge relocation where the heterogeneous catalyst plays a fundamental role.

\label{Concl}

\section{Acknowledgements}
\label{Acknow}

J.M. wants to acknowledge the support given by C\'atedras-CONACYT (Consejo Nacional de Ciencia y Tecnolog\'ia) under Project No. 1191; the support given by CONACYT through Project SEP- Ciencia B\'asica No. 156591; DGTIC (Direcci\'on General de C\'omputo y de Tecnolog\'ias de Informaci\'on y Comunicaci\'on) and the Supercomputing Department of Universidad Nacional Aut\'onoma de M\'exico for the computing resources under Project No. SC16-1-IR-29. C.D. wants to acknowledge the MSc. Scholarship provided by CONACYT with No.396279. The authors would also like to thank Dr. Adriana Longoria for helpful discussion.


 \section{References}
 \label{Ref}


\bibliographystyle{elec_comm_MS} 
\bibliography{gold_gen}







\pagebreak

\begin{table}
\begin{center}
\caption{Adsorption energies for each of the reactants and products along the reaction trajectory of triacetin transesterification.The values in parenthesis correspond to energies calculated without van der Waals contributions. $\Delta_{Diff}$ corresponds to the difference between values obtained with and without vdW effects}
\label{Table_Ads_Ener}
\begin{tabular}{p{1.0cm} p{3.0cm} p{2.0cm} p{3.0cm} p{2.0cm} }
\hline
\small{Step} & \small{Reactants energy (kcal/mol)} & \small{$\Delta_{Diff}$}  & \small{Products energy (kcal/mol)} & \small{$\Delta_{Diff}$} \\
\hline
1 & -62.59(-28.49) & -34.10 & -64.81(-28.34) & -36.47  \\
2 & -95.39(-53.06) & -42.33 & -99.84(-55.54) & -44.30 \\
3 & -82.47(-56.31) & -26.16 & -87.27(-56.44) & -30.83 \\
\hline
\end{tabular}
\end{center}
\end{table}


\begin{table*}
\begin{small}
\begin{center}
\caption{Main structural parameters of methyl acetate and diacetin with respect to the Au(111) surface at the Transition State TS2 of the first step in the overall reaction}
\label{Table_ts2_step1}
\vfill
\begin{tabular}{p{1.5cm} p{2.0cm} p{3cm} p{3cm}}
\hline
No. & Atoms		& Dist. (\AA) & Var. (\AA)\\
\hline
\multicolumn{4}{l}{Methyl acetate to Au(111) distance.}		\\
01 & Au42-031	 & 2.41  	& 0.33 	\\
\multicolumn{4}{l}{Diacetin to Au(111) distance.}		\\
02 & Au39-024	 & 2.44  	& -0.73	\\
03 & Au38-C25	 & 3.25  	& -0.62	\\
04 & Au35-C19	 & 3.47	 	& 0.03	\\
05 & Au50-C3	 & 4.60	 	& 0.75	\\
06 & Au47-O14	 & 4.87	 	& 1.19	\\
\hline
\end{tabular}
\end{center}
\end{small}
\end{table*}

\end{document}